\documentclass[prl,twocolumn]{revtex4-1}

\usepackage{amsmath,amssymb,graphicx,color}

\usepackage[pdftex,colorlinks=true,bookmarks=false,citecolor=blue,urlcolor=blue]{hyperref} %pdflatex

\begin{document}

\title{Enhanced nonlinear refractive index in epsilon-near-zero materials}

\author{L. Caspani$^1$, R. P. M. Kaipurath$^1$,  M. Clerici$^{1,2}$,   M. Ferrera$^1$, T. Roger$^1$,   J. Kim$^{3}$, N. Kinsey$^{3}$, M. Pietrzyk$^{4}$, A. Di Falco$^{4}$, V. M. Shalaev$^{3}$, A. Boltasseva$^{3}$, D. Faccio$^{1}$ 
}
\email{d.faccio@hw.ac.uk}
\address{$^1$Institute of Photonics and Quantum Sciences, SUPA, Heriot-Watt University, Edinburgh EH14 4AS, UK\\ 
$^2$School of Engineering, University of Glasgow, Glasgow G12 8LT, UK\\ 
$^3$School of Electrical and Computer Engineering and Birck Nanotechnology Center, Purdue University, 1205 West
State Street, West Lafayette, Indiana, 47907-2057 USA \\
$^4$SUPA, School of Physics and Astronomy, University of  St. Andrews, St. Andrews KY16 9SS, UK
}

\begin{abstract}
{New propagation regimes for light  arise from the ability to tune the dielectric permittivity to extremely low values.  Here we demonstrate a universal approach based on the low linear permittivity values attained in the epsilon-near-zero (ENZ) regime for enhancing the nonlinear refractive index, which enables remarkable light-induced changes of the material properties. Experiments performed   on Al-doped ZnO (AZO) thin films  show a six-fold increase of the Kerr nonlinear refractive index ($n_2$)  at the ENZ wavelength, located in the 1300 nm region. This in turn leads to ultrafast light-induced refractive index changes of the order of unity, thus representing a new paradigm for nonlinear optics.}
\end{abstract}

\maketitle
The nonlinear optical response of matter to light is, by its very nature, a perturbative and hence typically weak effect. Applications, e.g. for nonlinear optical switches or quantum optics, are therefore largely underpinned by the continuous endeavour to attain stronger and more efficient light-matter interactions. \\
Nonlinear mechanisms can typically be classified as resonant or non-resonant, depending on the frequency of light with respect to the characteristic electronic resonances of the material. Non resonant nonlinearities, like those present in transparent crystals or amorphous materials (e.g. fused silica glass), are generally weak and require high light intensities and/or very long samples to take advantage of an extended light matter interaction. Conversely, resonant nonlinearities can be several orders of magnitude stronger, but this comes at the price of introducing detrimental losses. A typical example is that of metals, which both reflect and absorb light strongly \cite{flytz,owens,boyd_gold}.
An alternative approach to enhance the nonlinear response of a material consists of creating artificial electromagnetic resonances, for example by stacking materials of different refractive index or using other types of composite materials \cite{smol,renger,stockman,bouh,kim,ko,Boyd:1994vp,vlad_book}.  
Creating resonant metal-dielectric stacks and composites yields a very strong nonlinear enhancement \cite{boyd_PC,boyd3,baumberg}, but inevitably exacerbates the detrimental role of linear and nonlinear losses.\\
Here we propose a different approach to enhance the effective nonlinearity without resorting to optical resonances. Our approach relies on enhancing the nonlinear effect, measured in terms of the nonlinear Kerr index $n_2$, rather than on a direct enhancement of the intrinsic $\chi^{(3)}$ nonlinear susceptibility.
As we show below, this enhancement arises due to the fact that the nonlinear refractive index is a function of both the nonlinear susceptibility and the linear refractive index. Recent progress in material design and fabrication has provided access to the full range of linear optical properties bounded by dielectric and metallic regimes. Of particular relevance for this work are materials which exhibit a real part of the dielectric permittivity that is zero, or close to zero, such as transparent conducting oxides where their permittivity cross over is typically located in the near infrared spectral region.\\
The linear properties of these  ``epsilon-near-zero''  (ENZ) materials have been investigated \cite{alu1,alu2,eps0_1,eps0_2,eps0_3,eps0_4,eps0_5,eps0_6,eps0_7,eps0_8,eps0_9,eps0_11,eps0_12,eps0_15,eps0_14,PRB_perfect_absorption_Sinclair,SciRep_perfect_absorption_Hwangbo,SciRep_perfect_absorption_and_switching_Hwangbo} for applications ranging from controlling the radiation pattern of electromagnetic sources to novel waveguiding regimes and perfect absorption. Similarly, the nonlinear properties have also been shown to be largely effected by the ENZ condition \cite{eps0_13,ciattoni:2010,ciattoni2,PRB2013_Scalora,ciattoni3,Ciattoni:10,zayats,Rizza_11}, and recently it has been theoretically predicted that the interplay between linear and nonlinear properties of ENZ bulk materials may allow three-dimensional self-trapping of light \cite{SciRepMarini2016}. However, experimental evidence  reported so far is limited to phase matching-free conditions in four-wave-mixing \cite{Suchowski:2013fa}, enhanced third and second harmonic generation \cite{THG_ENZ_ITO,APL_THG_CampioneScalora,SHG_ENZ_ITO}, and ultrafast optical switching \cite{AZO_optica}.\\
\indent In order to illustrate how the nonlinear Kerr index may be enhanced as a result of the ENZ {\emph{linear}} properties, we employed a  900 nm thick film of oxygen-deprived aluminium-doped zinc oxide (AZO) \cite{Azo1,AZO_optica}. The  AZO 900-nm thick thin film {was deposited by pulsed laser deposition \cite{Fab1,Fab2} (PVD Products Inc.) using a KrF excimer laser (Lambda Physik GmbH) operating at a wavelength of 248 nm for source material ablation (see Ref.~\cite{Azo1} for more details). \\
\indent The linear response i.e. real and imaginary parts of $\varepsilon$, $\varepsilon_r$ and $\varepsilon_i$, respectively, were measured by a standard reflection/transmission measurement using a tunable-wavelength, 100 fs, 100 Hz repetition rate, weak probe beam, see Figs.~\ref{fig1_linear}(a) and (b). The  linear permittivity, shown in Fig.~\ref{fig1_linear}(c) is then evaluated from the reflection/transmission measurements by means of an inverse transfer matrix approach: this allows to evaluate the complex permittivity from the measured reflectivity and transmissivity. The condition $\varepsilon_r=0$ (ENZ wavelength) is at $\sim$1300 nm.
\begin{figure}
\includegraphics[width=8.6cm]{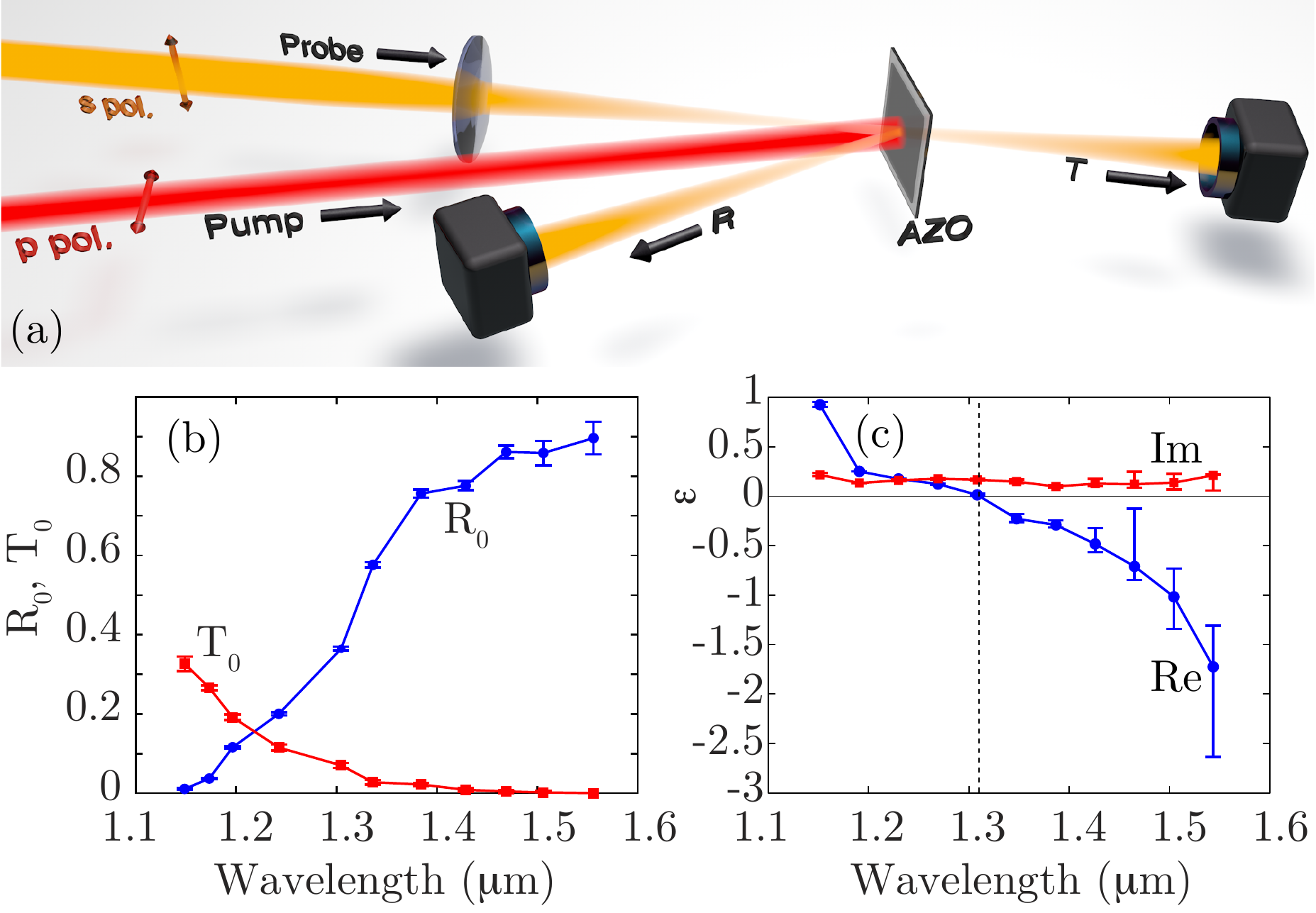}
\caption{(a) Experimental setup. A high intensity, horizontally polarised beam at 785 nm pumps the thin AZO film at normal incidence. The reflection and transmission of a weak probe beam (variable wavelength between 1150 and 1550 nm, vertically polarised, $\sim10^\circ$ angle of incidence) are simultaneously recorded. For the linear characterisation the pump beam was blocked. (b) Measured reflectivity ($R_0$, blue circles) and transmissivity ($T_0$, red squares)  in the linear regime (no pump beam). The error bars have been evaluated as the standard deviation on a sample of 300 measurements. (c) Real (blue circles) and imaginary (red squares) part of the linear permittivity extracted from the data in (a) using an inverse transfer matrix approach. The error bars have been evaluated by extracting $\epsilon$ from the pairs $\{R_0+\sigma_{R_0},T_0+\sigma_{T_0}\}$ (upper bounds) and $\{R_0-\sigma_{R_0},T_0-\sigma_{T_0}\}$ (lower bounds), where $\sigma_{T_0}$ and $\sigma_{R_0}$ are the reflectivity and transmissivity standard deviations, respectively. The large error bars for the real part of the permittivity at longer wavelengths are due to the very low values of transmissivity measured, in turn resulting in a high relative error.\label{fig1_linear}}
\end{figure}\\
For many applications and measurements, the nonlinear Kerr index $n_2$ is used instead of the third-order nonlinear $\chi^{(3)}$ tensor. For the case of a non-degenerate pump-probe scenario with a weak probe and an intense pump beam, the nonlinear index is given by \cite{boyd} 
\begin{equation}\label{eq:n2_complex}
n_2=\frac{3}{2\varepsilon_0 c}\frac{\chi^{(3)}}{n_r^{\textrm{pump}}(n_r+in_i)}\,,
\end{equation}
where both $n_2$ and $\chi^{(3)}$ are complex quantities, $\epsilon_0$ is the vacuum permittivity, $c$ is the speed of light in vacuum, $n_{r,i}$ are the real and imaginary parts of the linear refractive index at the weak probe wavelength, $n_r^{\textrm{pump}}$ is the real part of the refractive index at the pump wavelength. The pump-induced nonlinear refractive index change is then given by $\delta n = n_2 I$, where $I$ is the intensity of the optical beam \cite{boyd}.\\
\indent We note that although it is generally desirable to minimise the absorption losses, the imaginary part of the material's linear response also plays a crucial role in determining the effective nonlinear response. This can be appreciated by separating the complex nonlinear Kerr index into its real and imaginary parts \cite{boyd2}:\\
\begin{eqnarray}\label{eq:n2}
n_{2r}&=&\cfrac{3}{2\varepsilon_0 c}\cfrac{n_r\chi^{(3)}_r+n_i\chi^{(3)}_i}{D} \label{eq:n2a}\\ 
n_{2i}&=&\cfrac{3}{2\varepsilon_0 c}\cfrac{n_r\chi^{(3)}_i-n_i\chi^{(3)}_r}{D} \label{eq:n2b}
\end{eqnarray}
where $D=n_r^{\textrm{pump}}(n_r^2+n_i^2)$. The imaginary part $n_{2i}$ is usually associated with what is known as the nonlinear absorption coefficient, $\beta_2=4\pi n_{2i}/\lambda$, where $\lambda$ is the vacuum wavelength.
Consequently, we see that it is the interplay between linear $(n_r,n_i)$ and nonlinear $(\chi^{(3)}_r,\chi^{(3)}_i)$ properties that defines the nonlinear index, and provides a means to  enhance or tailor the effective $n_{2r}$ and $\beta_2$ coefficients as a function of wavelength.\\
An insight on the underlying physical mechanism at play can be gained with the simplified model $\chi^{(3)}_r=\chi^{(3)}_i=constant$, which allows us to predict the $n_2$ and $\beta_2$ behaviour based only on the linear material properties. In this case, the wavelength dependence of $n_{2r}$ is  determined by the term $(n_r+n_i)/D$, whereas the nonlinear absorption coefficient is determined by $(n_r-n_i)/D$. Both these quantities are plotted in Fig.~\ref{fig2_theo}(a) and (b), respectively, starting from the measured frequency-dependent linear refractive index of our sample.
\begin{figure}[!tb]
\includegraphics[width=8.6cm]{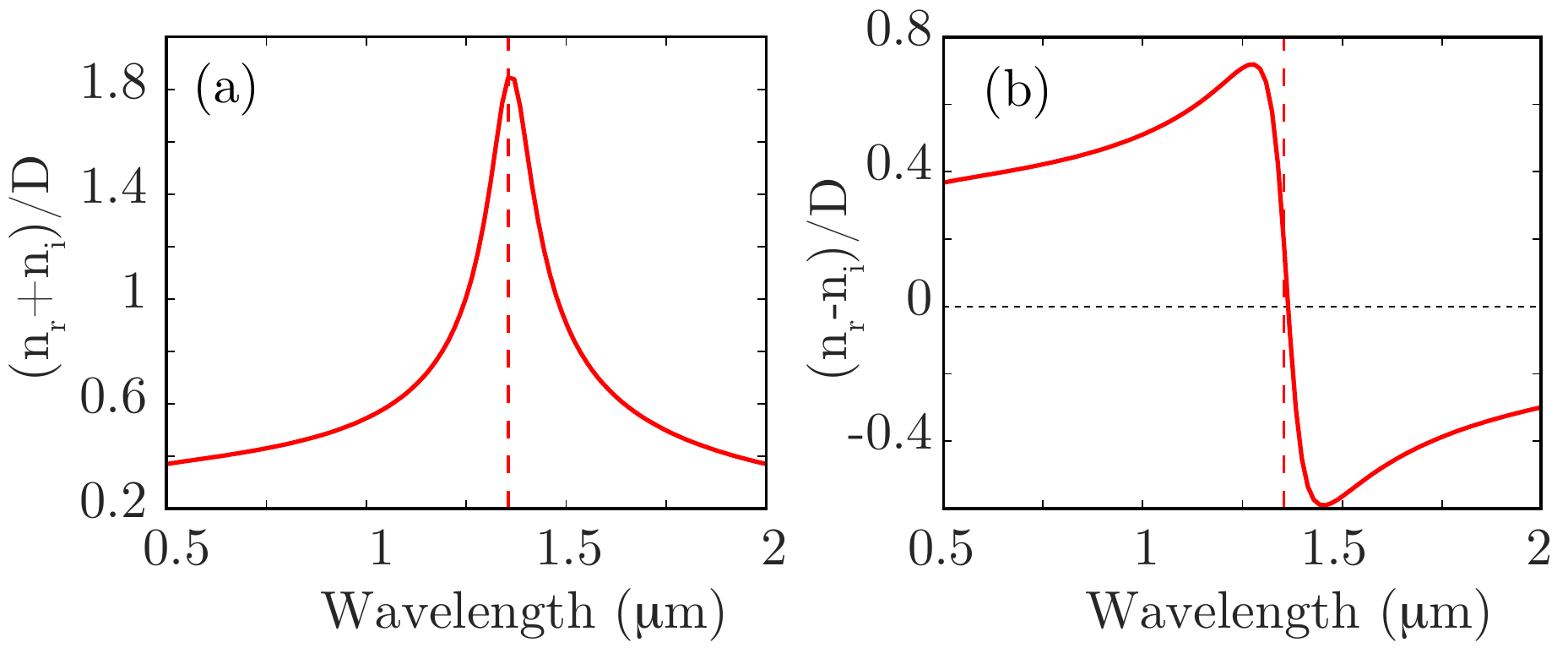}
\caption{Plots of the theoretically estimated trends for $n_2$ and $\beta_2$ from Eqs.~\eqref{eq:n2a} and \eqref{eq:n2b}.  (a) Plot of $(n_r+n_i)/D$ with $D=n_r^{\textrm{pump}}(n_r^2+n_i^2)$. This term weighs the real part of the nonlinearity, $n_{2r}$. (b) Plot of $(n_r-n_i)/D$: this term weighs the imaginary part of the nonlinearity, $\beta_{2}$. The vertical dashed lines indicate the ENZ wavelength \label{fig2_theo}}
\end{figure}\\
The presence of an ENZ wavelength (indicated by the vertical dashed line) significantly modifies the behaviour of the materials: a peak is observed in the $(n_r+n_i)/D$ term, indicating an enhancement of $n_{2r}$ whilst the nonlinear absorption shows a more complex behaviour and changes sign. 
The transition from positive to negative (i.e. saturable) nonlinear absorption occurs at the ENZ wavelength, such that the maximum nonlinear phase shift can be attained with zero nonlinear losses. This is a unique feature of ENZ materials and underlines the crucial role played by the material's linear dispersion and the effects of the ENZ condition on the nonlinear optical response. As discussed below, the condition $\chi^{(3)}_r=\chi^{(3)}_i$ is not strictly required for observing the $n_2$ enhancement at the ENZ condition.\\
We note that in order to observe the described enhancement one needs to achieve the ENZ condition for the real part of the linear dielectric permittivity whilst maintaining a relatively low imaginary part, as this in turn guarantees a low real part of the refractive index. 
The AZO film employed here displays all of the required properties for both the linear and nonlinear susceptibility to experimentally demonstrate the predicted nonlinear enhancement. Remarkably, this also comes with an extremely high damage threshold -- no damage was observed up to 2 TW/cm$^2$ (at 785 nm, 100 Hz repetition rate), to be compared to the few GW/cm$^2$ typical of metallic structures. \\
We characterised the nonlinear response of the AZO film with a pump and probe system [Fig.~\ref{fig1_linear}(a)] by measuring the pump-induced change in reflectivity and transmissivity for different pump intensities, see Figs.~\ref{fig3_RT_nonlinear}(a) and (b).
\begin{figure}
\begin{center}
\includegraphics[width=8.6cm]{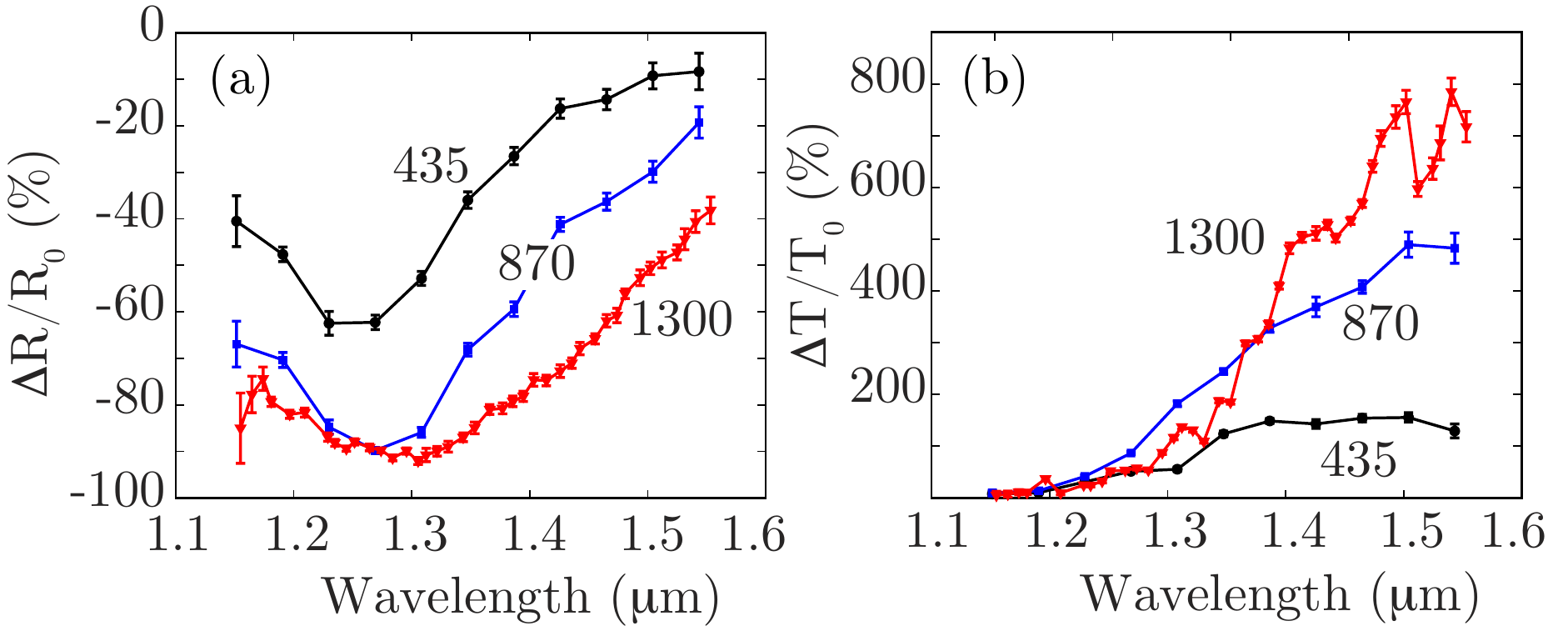}
\caption{(a) Measured reflectivity change $\Delta R/R_0=(R-R_0)/R_0$, where R is the reflectivity at the probe wavelength with the pump and $R_0$ is the linear value (pump off), for increasing pump intensities: $I_p=435$ GW/cm$^2$ (black circles), $I_p=870$ GW/cm$^2$ (blue squares), and $I_p=1300$ GW/cm$^2$ (red triangles). (b) Same as (a) but for transmissivity change $\Delta T/T_0=(T-T_0)/T_0$. The error bars have been evaluated propagating the error of the $R$, $R_0$, $T$, and $T_0$ standard deviations, from a sample of 300 measurements.\label{fig3_RT_nonlinear}}
\end{center}
\end{figure}
We observe a large variation in the reflectivity around the ENZ wavelength. Conversely,  the highest values of relative transmissivity change are observed for longer wavelengths, which is simply due to the normalisation with respect to the initial transmission $T_0$ that is very close to zero in this spectral region.\\
The pump pulse (100 fs, 100 Hz repetition rate, horizontally polarised) has a fixed wavelength $\lambda_{\rm pump}=785$ nm and is at normal incidence with respect to the sample. The probe pulse (100 fs, 100 Hz repetition rate, vertically polarised}) from an Optical Parametric Amplifier (OPA) and tunable from 1150 to 1550 nm is incident at a small ($<10^\circ$) angle. The intensity of the probe beam is kept low (below the GW/cm$^2$ level) to avoid any nonlinearity from the probe itself (no change in the probe transmissivity and reflectivity was observed at this intensity). The probe beam waist ($w_{0,{\rm probe}}=45$ $\mu$m) is  smaller than the pump beam waist ($w_{0,{\rm pump}}=125$ $\mu$m), in order to obtain a uniform pump intensity across the probe beam. For all the measurements the pump beam size was constant and the intensity was changed only increasing the energy. The pump-probe delay was then optimised to maximise the nonlinear effect.\\
For each probe wavelength we measure the pump-induced change in reflection and transmission of the probe beam and then use these values to retrieve the permittivity in the pumped case ($\varepsilon_{\rm nl}$) by applying  an inverse transfer matrix approach.
Whenever the dependence of $\varepsilon_{nl}$  from $I_p$ is linear, we may determine the third order nonlinearity from the relation \cite{SciRep_ag}:
\begin{equation}
\chi^{(3)}(\omega_p,\omega_{\rm probe}) = \frac{n_r^{\textrm{pump}}\varepsilon_0 c}{3}\frac{\partial \varepsilon_{nl}(\omega_{\rm probe},I_p)}{\partial I_p}\,,
\end{equation}
where $\omega_p$ and $I_p$ are the pump frequency and intensity, respectively, and the derivative can be evaluated as the slope of the linear fit of $\varepsilon_{\rm nl}(I_p)$. In Fig.~\ref{fig4_chi3}(a) we show an example of $\varepsilon_{\rm nl}(I_p)$ for a specific probe wavelength ($\lambda_{probe}=1258$ nm) with the corresponding linear fit, while Fig.~\ref{fig4_chi3}(b) reports the resulting real and imaginary parts of the material $\chi^{(3)}$ as a function of the probe wavelength.
\begin{figure}
\begin{center}
\includegraphics[width=8.6cm]{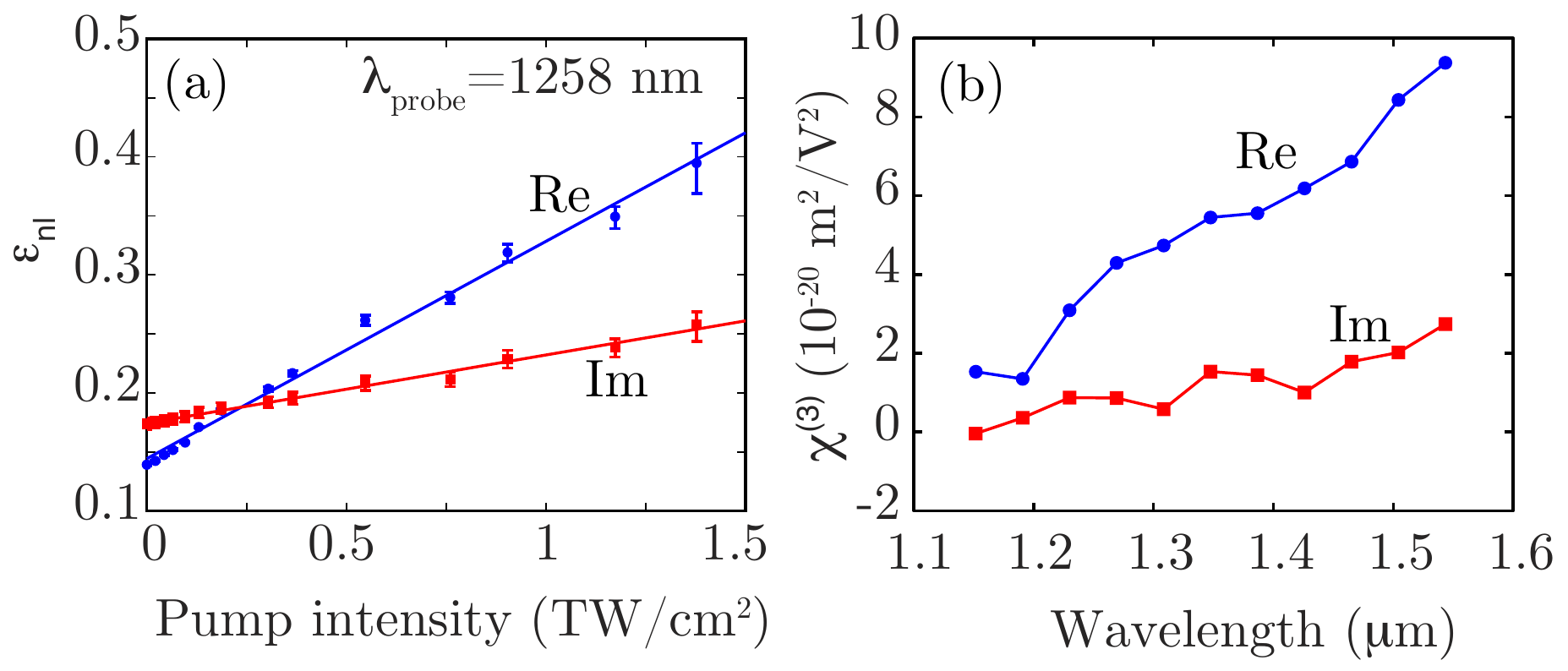}
\caption{(a) Real (blue circles) and imaginary (red squares) part of $\varepsilon_{\rm nl}$ as a function of the pump intensity for a specific probe wavelength, and corresponding linear fits (solid lines). (b) Real (blue circles) and imaginary (red squares) part of the $\chi^{(3)}$ tensor for different probe wavelengths.\label{fig4_chi3}}
\end{center}
\end{figure}\\
From the $\chi^{(3)}$  values it is possible to extract the nonlinear Kerr index by exploiting the formulas in Eqs.~\ref{eq:n2}. The results for both the real part, $n_{2r}$, and nonlinear absorption coefficient, $\beta_2$, are presented in Figs.~\ref{fig5_n2b2}(a) and (b), and show a good qualitative agreement with the theoretical curves in Fig.~\ref{fig2_theo}.
Most importantly, as seen in Fig.~\ref{fig5_n2b2}(a), a clear six-fold enhancement of $n_{2,r}$ (with respect to its lowest value at 1152 nm) is observed around the ENZ wavelength.
\begin{figure}[!t]
\begin{center}
\includegraphics[width=8.6cm]{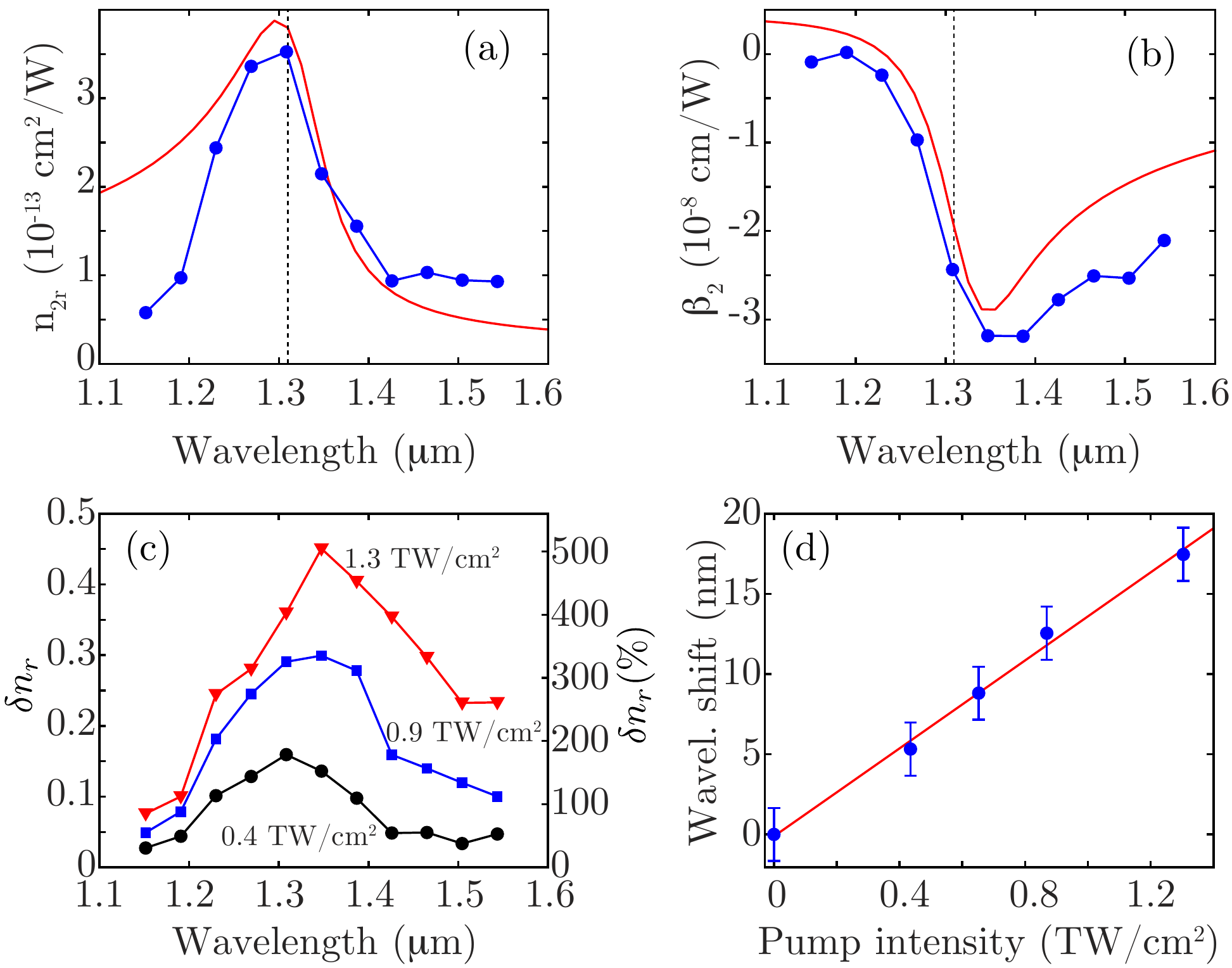}
\caption{(a) Real part of the nonlinear Kerr index, $n_{2r}$, and (b) nonlinear absorption coefficient, $\beta_2=4\pi n_{2i}/\lambda$, obtained from Eq.~\eqref{eq:n2a} and the data in Fig.~\ref{fig4_chi3} (blue dots). The red lines represent the expected theoretical result obtained from Eq.~(\ref{eq:b2_fit}). (c) Change in the real part of the refractive index, $\delta n_r=n_r(I_p)-n_r(I_p=0)$, for different pump intensities (left axis). Maximum relative refractive index change, $\delta n_r/n_r(I_p=0)$, at 1390  nm (right axis). (d) Measured shift of the probe carrier wavelength as a function of the pump intensity (the red curve is the linear fit).\label{fig5_n2b2}}
\end{center}
\end{figure}\\
To further support our analysis, we compare the theoretical predictions with the experimental results in Fig.~\ref{fig5_n2b2}. The red lines in Figs.~\ref{fig5_n2b2} (a) and (b) show the expected values of $n_2r$ and $\beta_2$ obtained from the measured linear refractive index and assuming a non dispersive $\chi^{(3)}$ with comparable real and imaginary parts. In detail, we plot the quantities:
\begin{eqnarray}\label{eq:n2_beta2_fit}
n_{2r,theo}&=&\cfrac{3}{2\varepsilon_0 c}\cfrac{n_r A +n_i B}{n_r^{\textrm{pump}}(n_r^2+n_i^2)}\,\label{eq:n2r_fit}\\ 
\beta_{2,theo}&=&\cfrac{4\pi}{\lambda}\cfrac{3}{2\varepsilon_0 c}\cfrac{n_r B-n_i A}{n_r^{\textrm{pump}}(n_r^2+n_i^2)}\,\label{eq:b2_fit}
\end{eqnarray}
where $A$ and $B$ (representing $\chi^{(3)}_r$ and  $\chi^{(3)}_i$, respectively) are used as fitting parameters and $A=4\times10^{-20}$ m$^2$/V$^2$ and $B=1\times10^{-20}$ m$^2$/V$^2$. The relatively good agreement with the data shows that indeed our measurements are compatible with the assumption used in Fig.~\ref{fig2_theo}, i.e. $\chi^{(3)}_r\sim\chi^{(3)}_i={\rm constant}$.\\
Remarkably, for the laser pulse intensities used in our experiments the measured nonlinear refractive index $n_{2r}$ gives a change of refractive index in the medium ($\delta n_r=n_{2r}I_p$) that is of the order of the linear refractive index.\\
For example, at  1390 nm we measured a change of refractive index as high as $\delta n_r=0.4$ to be compared with the linear index $n_r=0.09$, recorded for the highest intensity $I_p$ = 1300 GW/cm$^2$ without observing any optical damage, see Fig.~\ref{fig5_n2b2}(c). 
This large modulation places the ENZ nonlinearity in AZO in a regime where the approximation of expanding the material polarisation in a power series breaks down \cite{boyd}.\\
We note that similar results are in principle expected in any medium displaying similar linear properties together with a weak $\chi^{(3)}$ dispersion and for $\chi^{(3)}_r\sim\chi^{(3)}_i$.  Most importantly, the ENZ condition is often achieved together with significant losses while AZO films are featured by both the ENZ condition combined with a relatively low imaginary part of the permittivity, $\varepsilon_i$. The latter condition ensures that the linear refractive index is significantly close to zero, which as discussed above maximises the observed enhancement of the nonlinear index.\\
Finally, in Fig.~\ref{fig5_n2b2}(d)  we show how the carrier wavelength of the probe pulse transmitted through the sample increases linearly with the pump intensity, and shifts up to $17.5\pm1.6$ nm, i.e. by more than the 15 nm probe input bandwidth. This ``nanoscale wavelength shifter'' could be applied e.g. for single photon wavelength division multiplexing \cite{Matsuda}. In our demonstration, large frequency shifts were achieved with high pump intensities (TW/cm$^2$), which might however be reduced by relying on nanostructured materials \cite{Munsken2014}.\\
In conclusion, ENZ materials allow one to tailor and access novel linear propagation regimes. Here we have shown the ability to exploit the ENZ regime for enhancing third-order nonlinear effects thus leading to an ultrafast light-induced metal-to-dielectric phase change. The interplay between the real and imaginary parts of the linear refractive index and $\chi^{(3)}$ tensor also leads to a peculiar wavelength-dependent behaviour of the nonlinear refractive index. This allows, for example, an enhancement of the real part of the nonlinear index, which in turn is associated with a nonlinear phase shift in the probe beam. On the other hand, novel and interesting behaviours are observed such as the change in sign of the $\beta_2$ coefficient, effectively eliminating nonlinear absorption close to the ENZ wavelength. Moreover, the possibility to optically control the material's refractive index by amounts comparable to the linear values (close to 500\% relative changes in the refractive index are reported here) may allow one to effectively tailor the impedance of the material and match it to that of the surrounding medium. The ability to access ultrafast light-induced refractive index changes of the order of unity represents a new paradigm for nonlinear optics.\\
During the review process, we became aware of a related study that has now been published \cite{Boyd_ITO}.
\section*{Acknowledgments}
L.C. and M.F. acknowledge the support from the People Programme (Marie Curie Actions) of the European Union's FP7 Programme under REA grant agreement no. 627478 (THREEPLE) and no. 329346 (ATOMIC), respectively. D.F. acknowledges financial support from the European Research Council under the European Union Seventh Framework Programme (FP/2007-2013)/ERC GA 306559 and EPSRC (UK, Grant EP/M009122/1). All data relevant to this work may be obtained at DOI: 10.17861/d82f29dc-5c47-4c5a-b893-c1ea93ab5224.
\end{document}